\documentclass[aps,pra,twocolumn]{revtex4}
\usepackage{amsmath}
\usepackage{graphicx}
\usepackage[section]{placeins}

\newcommand{\ket}[1]{|#1\rangle}
\newcommand{\bra}[1]{\langle#1|}

\newcommand{\eq}{\begin{equation}}
\newcommand{\fine}{\end{equation}}

\begin{document}

\title{Amplification of polarization NOON states}
\author{Chiara Vitelli$^{1}$,  Nicol\`{o} Spagnolo$%
^{1} $, Fabio Sciarrino$^{2,1}$, and  Francesco De Martini$^{1,3}$\\
$^{1}$Dipartimento di Fisica dell'Universit\'{a} ''La Sapienza'' and
Consorzio Nazionale Interuniversitario per le Scienze Fisiche della Materia,
Roma 00185, Italy\\
$^{2}$Centro di\ Studi e Ricerche ''Enrico Fermi'', Via Panisperna
89/A,Compendio del Viminale, Roma 00184, Italy\\
 $^{3}$Accademia Nazionale dei Lincei, Italy}

\begin{abstract}
NOON states are path entangled states  which can be exploited to
enhance phase resolution in interferometric measurements. In the
present paper we analyze the quantum states obtained by optical
parametric amplification of polarization NOON states. First we
study, theoretically and experimentally, the amplification of a
2-photon state by a collinear Quantum Injected Optical Parametric
Amplifier (QIOPA). We compared the stimulated emission regime with
the spontaneous one, studied by Sciarrino et al. (PRA \textbf{77},
012324), finding comparable visibilities between the two cases but
an enhancement of the signal in the stimulated case. As a second
step, we show that the collinear amplifier cannot be successfully
used for amplifying N-photon  states with N$>$2 due to the
intrinsic $\frac{\lambda }{4}$ oscillation pattern of the crystal.
To overcome this limitation, we propose to adopt a scheme for the
amplification of a generic state based on a non-collinear QIOPA
and we show that the state obtained by the amplification process
preserves the $\frac{\lambda}{N}$ feature and exhibits a high
resilience to losses. Furthermore, an asymptotic unity visibility
can be obtained when correlation functions with sufficiently high
order $M$ are analyzed.
\end{abstract}

\maketitle

\section{Introduction}

\begin{figure}[t!]
\centering
\includegraphics[scale=0.35]{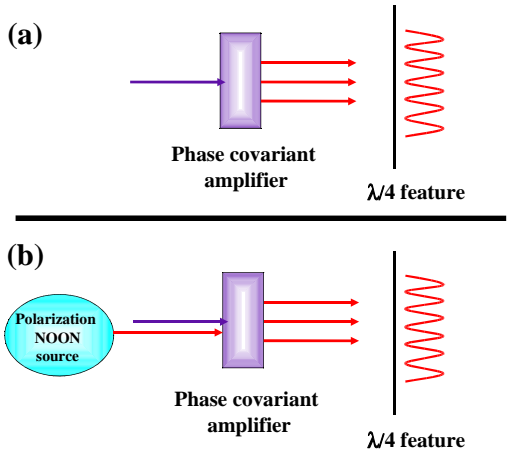}
\caption{ (a)Unseeded optical parametric amplifier
.\ (b) Amplification of a polarization entangled NOON
state.}
\label{fig:confronto}
\end{figure}

In the last few years it has been proposed to exploit quantum effects to provide resolution enhancement in imaging procedures. Among the numerous problems that are currently studied under the general name of \textsl{quantum imaging}, the investigations concerning the quantum limits on optical resolution have a special importance, as they may lead to new concepts in microscopy and optical data storage. Such so-called \textsl{super-resolution} techniques, studied for a long time at the classical level in the perspective of beating the Rayleigh limit of resolution, were recently revisited at the quantum level \cite{Boto00}. It was shown that it was possible to improve the performance of super-resolution techniques by adopting non-classical light \cite{Holl93,Giov06}. This approach,\textsl{ quantum lithography}, may lead in the future to innovative microscopy techniques, to record features in images which are much smaller than the wavelength of the light or to improve the optical storage capacity beyond the wavelength limit. 
In such framework, path entangled NOON states 
$\ket{\psi^{N}}_{AB}=\frac{1}{\sqrt{2}}\left( \left|
N\right\rangle _{A}\left| 0\right\rangle _{B}+\left|
0\right\rangle _{A}\left| N\right\rangle _{B}\right) $ have been adopted 
to increase the resolution in quantum interferometry. Indeed
in such states a single mode phase shift $\varphi$
induces a relative shift between the two components equal to
$N\varphi $. This feature leads to a sub-Rayleigh resolution scaling as
$\frac{\lambda }{2N}$, $\lambda$ being the wavelength of the field \cite{DAng01}.
The theorethical and experimental study of photonic NOON states \cite{Kapa06} has lead 
to the experimental a posteriori generation of two, three and four photons states
\cite{Edam02,Mitc04,Walt04} and to the conditional generation of a state with 
$N=2$ \cite{Eise05}.
Furthermore, very recently schemes for the generation of path-entangled NOON states
with high value of fidelity and arbitrary N have been proposed
\cite{Hofm07,Dowl07}. However, the weak value of the generated
number of photons strongly limits the potential applications to
quantum lithography and quantum metrology. Furthermore a NOON
state, as any superposition of macroscopic states,\ is
``supersensitive'' to losses.\ Hence for a N-photon  state a
fractional loss $\frac{1}{N}$\ would destroy the quantum effect
responsible for the phase resolution improvement.
\newline A natural approach to increase the number of
photons and to minimize the effect of losses is to exploit the
process of stimulated emission. This process, also known as
quantum injected optical parametric amplification, has been
first studied in \cite{DeMa98}, and has found some important applications
in the context of quantum information \cite{DeMa02,DeMa05_2,DeMa07}.
Recently the output radiation of an unseeded optical parametric amplifier (OPA) has
been exploited to show the typical $\lambda/4$ feature
\cite{DeMa08} Fig.\ref{fig:confronto}-(a).
In the present paper we investigate the task of the amplification of photonic
NOON states by two different devices, both based on a quantum injected optical
parametric amplifier (QIOPA). First, in Sec.\ref{sec:seed}, we review how a sub-Rayleigh
$\frac{\lambda}{2N}$ resolution can be obtained by an interferometric device acting on a NOON
state and we show how the performances of this scheme are affected by losses.
Then, in Sec.\ref{sec:exp} we study both theoretically and experimentally the
amplification of a 2 photon state by a collinear QIOPA, as shown schematically
in Fig.\ref{fig:confronto}-(b), investigating how the features of the state are
modified when the entanglement is broadcasted via amplification over a large number of particles.
An experimental comparison with the spontaneous field
Fig.\ref{fig:confronto}-(a) of the collinear OPA, that  intrinsically
has a $\frac{\lambda}{4}$ feature \cite{DeMa08}, shows that the
visibilities in the two regimes are comparable, while the injected
case manifests an increase of the signal due to the stimulated
emission process. We then show that this device cannot be
successfully used to amplify a generic N-photon state since the
typical $\frac{\lambda}{2N}$ feature of the seed is lost. Finally,
in Sec.\ref{sec:non_coll}, we propose to exploit a non-collinear
QIOPA in order to amplify a generic state maintaining the
interference pattern of the seed, showing that significant value
of the visibilities can be achieved by investigating 
high-order correlation functions. Finally, the effects of losses
are investigated, demonstrating that the amplified field exhibits
a higher resilience to losses with respect to a pure NOON state.

\section{Interferometrical pattern of the seed and decoherence}
\label{sec:seed}
In this section we recall the interferometrical pattern of the injected seed and we study how
the state features are affected by losses.\\
We begin with the polarization entangled NOON state $\vert
\psi^{N}
\rangle_{1}=\frac{1}{\sqrt{2}}\left(\ket{N+}-\ket{N-}\right)_{1}$.
In this case the N-photon state is entangled in the polarization
degree of freedom and belongs to the spatial mode
$\mathbf{k}_{1}$. Introducing a phase shift $\varphi$ between the
two orthogonal polarizations, the state reads
$\ket{\psi^{N}_{\varphi}}_{1}=\frac{1}{\sqrt{2}}\left(\ket{N+}-e^{\imath
N \varphi} \ket{N-}\right)_{1}$, where $\ket{p\xi}$ refers to the
quantum state with $p$ photons polarized
$\overrightarrow{\pi}_{\xi}$. The M-th order correlation function
$\mathcal{G}_{seed}^{(M)} = \bra{\psi^{N}_{\varphi}} \hat{a}^{\dag
M}_{H} \hat{a}^{M}_{H} \ket{\psi^{N}_{\varphi}}$ reads, for M=N:

\begin{equation}
\mathcal{G}^{(N)}_{seed} = \frac{N!}{2^{N}} \left[1 + (-1)^{N+1} \cos(N \varphi) \right]
\end{equation}
while for $M<N$ all the functions  do not exhibit any oscillation
behaviour and have the  expression:

\begin{equation}
\mathcal{G}^{(M)}_{seed} = \frac{N!}{2^{M} (N-M)!}
\end{equation}

\begin{figure}[ht!]
    \centering
        \includegraphics[width=0.40\textwidth]{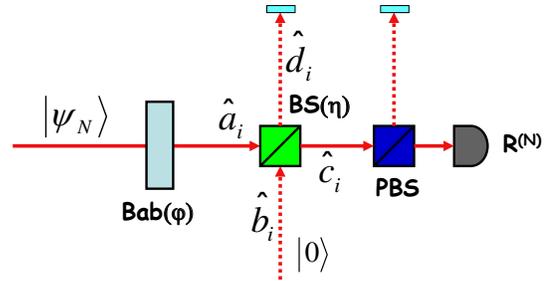}
    \caption{Decoherence model of the interferometric process. The phase shift in the NOON state $\vert \psi^{N} \rangle$ is inserted by a Babinet compensator, while the BS with efficiency $\eta$ models the decoherence process. The signal is then analyzed in polarization by the PBS and the N photons absorbing device $R^{(N)}$.}
    \label{fig:dec_seed_inj}
\end{figure}

In order to simulate losses in the transmission path and non
unitary detection efficiency, we now introduce a Beam Splitter
(BS) of transmittivity $\eta$,  as shown in
Fig.\ref{fig:dec_seed_inj}.

The density matrix after the decoherence process, obtained by the
insertion of the I/O BS relations and by tracing on the unrevealed
reflected mode, becomes:

\begin{equation}
\label{eq:rho_NOON_dec}
\hat{\rho}_{loss} = \eta^{N} \hat{\rho}_{NOON} + \sum_{i=0}^{N-1} \left( \begin{matrix} N \\ i \end{matrix}\right) \eta^{i} \left( 1 - \eta \right)^{N-i} \hat{\rho}_{i}
\end{equation}
where $\hat{\rho}_{NOON} = \ket{\psi^{N}_{\varphi}} \bra{\psi^{N}_{\varphi}}$ is
the density matrix of a pure NOON state, and
 $\hat{\rho}_{i} = \frac{1}{2} \left[ \ket{i+,0-} \bra{i+,0-} + \ket{0+,i-} \bra{0+,i-} \right]$
 is the density matrix of a mixed $i$ photons state. Only the first part of this quantum state contributes
 to the N-th order correlation function, and the successful events rate is reduced by a factor $\eta^{N}$.
  We finally obtain that the correlation function after losses reads:
\begin{equation}
\mathcal{G}^{(N)}_{loss} = \eta^{N} \mathcal{G}^{(N)}_{seed}
\end{equation}

We propose in the following sections to exploit an amplification
process in order to improve the robustness to losses of these states
without losing their $\frac{\lambda}{N}$ sub-Rayleigh feature.

\section{Collinear amplification of a 2 photon NOON state}
\label{sec:exp} In this section we study, both theoretically and
experimentally, the amplification of a two-photons
polarization-entangled NOON state exploiting an optical parametric
amplifier working in a collinear configuration. It will be shown
that this device cannot be used to amplify a generic N-photons
state as its $\frac{\lambda}{2N}$ oscillation pattern is masked by
the intrinsic oscillation of the amplification crystal.

\subsection{Theoretical approach}
As a first step we consider the generation of a two photon NOON
state  by spontaneous parametric down
conversion in a first crystal over the two polarization mode $%
\overrightarrow{\pi }_{+}$\ and $\overrightarrow{\pi }_{-}$,\ on the same
spatial mode $\mathbf{k_{1}}$.\ The state generated is $|\psi^{2} \rangle_{1} =\frac{%
1}{\sqrt{2}}(|2+\rangle -|2-\rangle )_{1}=|1H;1V\rangle_{1}$,
where $|p+;q-\rangle
$\ stands for the quantum state of $p$ photons polarized $\overrightarrow{%
\pi }_{+}$ and $q$ photons polarized $\overrightarrow{\pi }_{-}$.

The amplification of the  state $|\psi^{2} \rangle_{1} $ is
realized by injecting the quantum
state  into a QIOPA acting on the input field $\mathbf{k_{1}.%
}$\ The interaction Hamiltonian of the optical parametric amplification $%
\widehat{H}_{coll}=i\chi \hbar \widehat{a}_{1H}^{\dag }\widehat{a}%
_{1V}^{\dag }+h.c.$\ acts on the spatial mode $\mathbf{k_{1}}$.\ The output
state over the mode $\mathbf{k_{1}}$\ is: {\small
\begin{equation}
|\Phi ^{2}\rangle_{1} =\frac{1}{C}\sum_{n=0}^{\infty }\Gamma
^{n-1}\left(\frac{n}{C^{2}} -\Gamma^{2}\right) |nH;nV\rangle _{1}
\end{equation}
}with $C=\cosh g$,\ $\Gamma =\tanh g$,\ being $g$\ the non-linear
gain of the amplification process \cite{DeMa05,DeMa01}.\newline
The peculiar $\lambda /4$ interference path feature of a two
photon NOON state,\ can be investigated by performing an
interferometric measurement on the amplified
field.\ To this end a phase shift $\theta $ is introduced, after the amplification stage, in the $\{%
\overrightarrow{\pi }_{+},\overrightarrow{\pi }_{-}\}$ basis,\ corresponding
to a rotation of an angle $\theta /2$\ in the basis $\{\overrightarrow{\pi }%
_{H},\overrightarrow{\pi }_{V}\}$. The state is then analyzed in
polarization and detected adopting single photon detectors. The
amplified signal can be evaluated by the first order
correlation function $\mathcal{G}^{(1)}_{N=2}=\ _{1}\langle \Phi^{2} |\widehat{c}_{1}^{\dag }%
\widehat{c}_{1}|\Phi^{2} \rangle_{1} $,\ where
$\widehat{c}_{1}^{\dag }=\left( \cos \theta
/2\widehat{a}_{1H}^{\dag }-\sin \theta /2\widehat{a}_{1V}^{\dag
}\right) $\ is the transmitted mode of a polarizing beam splitter
(PBS).\ We find that $\mathcal{G}^{(1)}_{N=2}=3\overline{n}+1$,\
independently of the phase value $\theta ,$ with
$\overline{n}=\sinh ^{2}g$.\ The state generated by the amplifier
is then investigated through the second order
correlation function $\mathcal{G}^{(2)}_{N=2}=\ _{1}\langle \Phi^{2} |\widehat{c}_{1}^{\dag }%
\widehat{c}_{1}^{\dag }\widehat{c}_{1}\widehat{c}_{1}|\Phi^{2}
\rangle_{1} $.\ By tuning the phase shift $\theta $,\ we find that
the expression of the second order correlation function is:

\begin{equation}
\label{eq:G22_coll}
\mathcal{G}_{N=2}^{(2)}=2\overline{n}(4+7\overline{n})+\frac{1}{2}(7\overline{n}%
^{2}+7\overline{n}+1)(1-\cos (2\theta ))
\end{equation}

The corresponding visibility of the obtained fringe pattern is calculated accordingly to the general definition:
\begin{equation}
\mathcal{V}^{(M)}_{N} = \frac{\mathcal{G}^{(M)}_{N}(max) - \mathcal{G}^{(M)}_{N}(min)}{\mathcal{G}^{(M)}_{N}(max) + \mathcal{G}^{(M)}_{N}(min)}
\end{equation}
where $M$ is the order of the correlation and $N$ is the number
 of photon of the injected seed. In the case of eq.(\ref{eq:G22_coll}) the visibility reads:
\begin{equation}
\mathcal{V}_{N=2}^{(2)}=\frac{7\overline{n}^{2}+7\overline{n}+1}{35\overline{n}%
^{2}+23\overline{n}+1}
\end{equation}

\begin{figure}[b]
\centering
\includegraphics[width=0.4\textwidth]{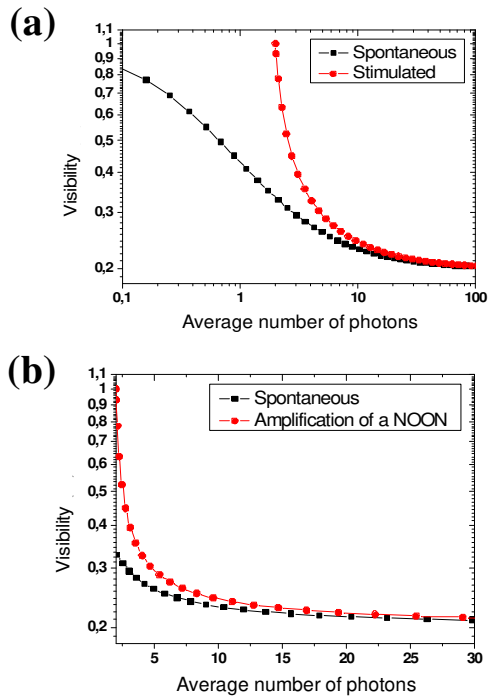}
 \caption{Theoretical trend of the visibility in
function of the number of photons generated by the amplification
in the two cases:\ spontaneous and stimulated.
}\label{fig:confronto_teorico_visi}
\end{figure}

We observe that a non-vanishing visibility is found for any value of $g:$ $%
\mathcal{V}_{N=2}^{(2)}(g\rightarrow \infty )=\frac{1}{5}$. The
fringe pattern
exhibits a dependence on $2\theta $ and hence a period equal to $\frac{%
\lambda }{2}$. This feature can be exploited to carry out
interferometry with sub-Rayleigh resolution, i.e., with fringe
period lower than $ \lambda $, in a higher flux regime compared to
the two photon configurations.\ The interest in amplifying a NOON
state belongs to the trend of visibility as a function of the
number of generated photons.  \ Recently, as said, it has been
demonstrated \cite{DeMa08} that the output field of a collinear
parametric amplifier working in spontaneous emission regime shows
a $\lambda/4$ feature. There an unseeded optical parametric
amplifier working in collinear regime was pumped by an UV beam.
The output radiation, after a phase shifter, was analyzed in
polarization. The fringe pattern visibility in that case was
$\mathcal{V}_{N=0}^{(2)}=\frac{\overline{n}+1}{5\overline{n}+1}$.
The asymptotical values of visibilities in the two regimes,
spontaneous and stimulated, are equal;\ on the contrary for an
intermediate number of generated photons the visibility in the
amplified regime is higher than that in the spontaneous one as
shown in Fig. \ref{fig:confronto_teorico_visi}-(a). Hence, the
injection of a seed with theoretical visibility equal to 1
leads to an advantage in the visibility for the amplified field
with respect to the case of spontaneous emission. \ We note that
the same average number of photons in the two regimes is achieved
for different values of the gain. In the spontaneous regime the
average photons number generated by the amplifier is
$\langle \hat{n} \rangle_{sp}=2\sinh^{2}g$, on the contrary, for the same gain
value, in the stimulated regime we have:
$\langle \hat{n}\rangle_{stim}=2+6\sinh^{2}g$.  For
a value of the gain $g=0$\  the number of photons in
the stimulated case is $\langle \hat{n} \rangle_{stim}=2$,\ unlike the
spontaneous case in which $\langle \hat{n} \rangle_{sp}=0$.\ In both cases the
value of visibility tends to $1$ for $g\rightarrow\;\;0$.\ By
analyzing the trends of visibilities in
Fig.\ref{fig:confronto_teorico_visi}-(b) we see that the advantage
of amplifying a NOON state holds until $\langle \hat{n} \rangle \simeq 30$.

\begin{figure}[tbph]
\centering
\includegraphics[width=0.4\textwidth]{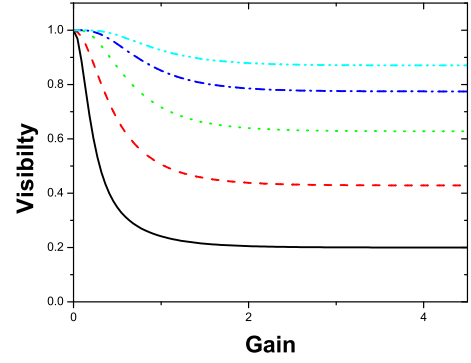}
\caption{Plot of the visibilities $\mathcal{V}^{(M)}_{N=2}$ with $2\leq
M\leq 6$ for the collinear QIOPA in stimulated emission with the
injection of a 2-photon NOON state  as a function of the nonlinear
gain $g$.
Straight line corresponds to $\mathcal{V}^{(2)}_{N=2}$, dashed line to $\mathcal{V}%
^{(3)}_{N=2}$, dotted line to $\mathcal{V}^{(4)}_{N=2}$, dash-dotted line do $\mathcal{V}%
^{(5)}_{N=2}$ and short dash-dotted line to $\mathcal{V}^{(6)}_{N=2}$.}
\label{fig:v_stim_N=2}
\end{figure}

An enhancement of the fringe pattern can be obtained by evaluating
the M-th order visibility $\mathcal{V}_{N=2}^{(M)}$,\ with $ M>2$,
corresponding to the M-th order correlation function at time t:
$\mathcal{G}_{N=2}^{(M)}=\ _{1}\langle \psi^{2}
|[\hat{c}_{1}^{\dag }(t)]^{M}[\hat{c}_{1}(t)]^{M}|\psi^{2}
\rangle_{1}$.  This calculation have been
performed in the Heisenberg picture, where the field operator
$\hat{c}^{\dag}_{1}(t)$ is the time evolution of the analyzed
field $\hat{c}_{1}^{\dag}$  solving the Heisenberg equations for
the collinear OPA. We calculated the first 6 orders  correlation
functions, obtaining the following visibilities:
\begin{eqnarray}
\mathcal{V}_{N=2}^{(2)} &=& \frac{1 + 7 \overline{n} + 7 \overline{n}^{2}}{1 + 25 \overline{n} + 35 \overline{n}^{2}} \\
\mathcal{V}_{N=2}^{(3)} &=& \frac{12 + 48 \overline{n} + 39 \overline{n}^{2}}{12 + 84 \overline{n} + 91 \overline{n}^{2}} \\
\mathcal{V}_{N=2}^{(4)} &=& \frac{12 + 291 \overline{n} + 822 \overline{n}^{2} + 567 \overline{n}^{3}}{12 + 291 \overline{n} + 1078 \overline{n}^{2} + 903 \overline{n}^{3}} \\
\mathcal{V}_{N=2}^{(5)} &=& \frac{135 + 1315 \overline{n} + 2845 \overline{n}^{2} + 1705 \overline{n}^{3}}{135 + 1315 \overline{n} + 3245 \overline{n}^{2} + 2201 \overline{n}^{3}} \\
\mathcal{V}_{N=2}^{(6)} &=& \frac{45 + 1745 \overline{n} + 10080 \overline{n}^{2} + 17507 \overline{n}^{3} + 9245 \overline{n}^{4}}{45 + 1745 \overline{n} + 10080 \overline{n}^{2} + 18657 \overline{n}^{3} + 10621 \overline{n}^{4}}
\end{eqnarray}
The theoretical plots of the visibilities are reported in Fig.\ref
{fig:v_stim_N=2}. We observe that an increasing trend is obtained
by exploiting correlation functions with higher order $M$. This
means that analyzing a higher order absorption process the contrast
of the fringe pattern is enhanced. This feature was also predicted
in the spontaneous emission regime in \cite{Agar07}, and
experimentally observed in \cite{DeMa08}.

\subsection{Experimental verification}
\begin{figure}[t]
\includegraphics[width=0.5\textwidth]{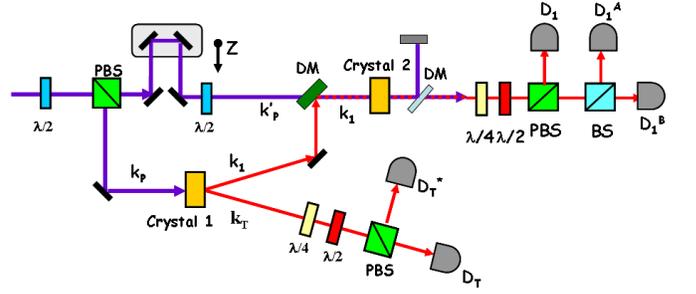}
\caption{Experimental scheme adopted to amplify a 2-photon
state.By measuring coincidences between detector
$\{D_{T},D_{T}^{\ast }\}$\ on spatial mode $\mathbf{k_{T},}$\ the
state on spatial mode $\mathbf{k_{1}}$\ is prepared in the
two-photon NOON state $|\psi^{2}\rangle_{1}$.The rate of the
trigger signal was around $10^{.}000Hz$\ and the rate of
coincidences between $(D_{T},D_{T}^{\ast })$\ was around $400Hz$.
} \label{fig:setup}
\end{figure}

The previous theoretical results have been experimentally verified
adopting an injected high-gain optical parametric amplifier.\ The
experimental setup is sketched in Fig \ref{fig:setup}.

The excitation source was a Ti:Sa Coherent MIRA mode-locked laser
amplified by a Ti:Sa regenerative REGA device operating with pulse
duration $180$\textrm{fs} at a repetition rate of
$250$\textrm{kHz}. The output beam, frequency-doubled by second
harmonic generation, provided the excitation beam of UV wavelength
(wl) $\protect\lambda _{P}=397.5$\textrm{nm} and power
$750$\textrm{mW}. The UV  beam was split in two beams through a
$\lambda /2$ waveplate and a polarizing beam splitter (PBS) and
excited two BBO ($\beta $-barium borate) NL crystals cut for type
II phase-matching.  The pump power of beam $\mathbf{k}_{P}$ was
set in order to have a negligible probability to generate three
couples of photons ($<10\%$).  Let us describe how the 2-photon
state $\ket{\psi^{2}}_{1}=2^{-1/2}\left( |2+\rangle -|2-\rangle
\right)_{1} =|1H;1V\rangle_{1} $\ was conditionally generated on
mode $\mathbf{k}_{1}$.\ We adopted the scheme demonstrated by
Eisenberg et al \cite{Eise04}: Crystal 1, excited by the beam
$\mathbf{k}_{P} $, is the spontaneous parametric down-conversion
(SPDC) source of entangled photons of wavelength $\lambda
=2\lambda _{P}$, emitted over the two output modes
$\mathbf{k}_{i}$ ($i=1,T$),\ where $T$\ stands for the trigger
mode,\ in the state $|\Psi _{2}^{-}\rangle_{1T}
=\frac{1}{\sqrt{3}}(|2H\rangle _{1}|2V\rangle _{T}-|1H;1V\rangle
_{1}|1H;1V\rangle _{T}+|2V\rangle _{1}|2H\rangle _{T})$. The two
photons associated to mode $\mathbf{k}_{T}$ were coupled into a
single mode fiber and excited two single photon counting module
(SPCM) $\left\{D_{T},D_{T}^{\ast} \right\}$.\ The state $|1H;1V\rangle _{T}$\ was
detected on mode $\mathbf{k_{T}}$\ by measuring the coincidences
between detectors $\{D_{T},D_{T}^{\ast }\}$\ in the $\{\overrightarrow{\pi }_{H},%
\overrightarrow{\pi }_{V}\}$ polarization basis on mode
$\mathbf{k}_{T}$ leading to the conditional preparation of the
state $|1H;1V\rangle _{1}$ on mode $\mathbf{k}_{1}$.

\begin{figure}[t]
\centering
\includegraphics[width=0.50\textwidth]{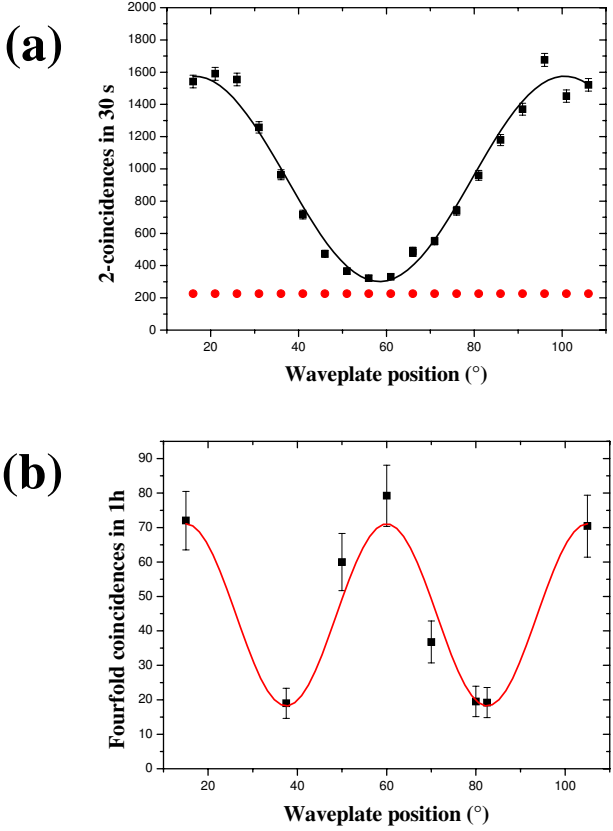}
\caption{(a)Fringe pattern of the two-fold-coincidences between
detectors $\{D_{1}^{B},D_{T}\}$ (b) Fringe pattern of the
four-fold-coincidences between detectors
$\{D_{1},D_{1}^{B},D_{T},D_{T}^{\ast}\}$.}
\label{fig:injected_NOON}
\end{figure}

The amplification of the injected 2-photon
state was achieved by superimposing the pump beam on mode $\mathbf{%
k_{p}^{^{\prime }}}$ and the field on mode $\mathbf{k_{1}}$ on
crystal II exploiting a dichroic mirror ($DM$) with high
reflectivity at $\lambda $ and high transmittivity at $\lambda_{p}
$. The output radiation was then analyzed through a polarizing
beam splitter (\textrm{PBS}) and detected
adopting single photon detectors SPCM-AQR14 ($\mathrm{D}_{1}^{A},\mathrm{D}%
_{1}^{B},\mathrm{D}_{1}$).\\ In order to characterize the state
produced by the first crystal, a measurement of the second order
correlation function of the injected field,\ without the
contribution of the UV pump beam on crystal 2,\ was carried out.\
The typical $\lambda /4$\ fringe pattern was measured by the
fourfold coincidences between detectors
$\{D_{1},D_{1}^{B},D_{T},D_{T}^{\ast }\}$, through evaluation of
the second order correlation function $G_{seed}^{(2)}=\
_{1}\bra{\psi^{2}}\widehat{c}_{1}^{\dag}\widehat{c}_{2}^{\dag}\widehat{c}_{2}\widehat{c}_{1}\ket{\psi^{2}}_{1}$,\
where $\widehat{c}_{2}^{\dag}=\left(\sin\theta/2
\widehat{a}_{1H}^{\dag}+\cos\theta/2
\widehat{a}_{1V}^{\dag}\right)=\widehat{c}_{1^{\perp}}^{\dag}$.\
The obtained visibility $\mathcal{V}^{(2)}_{seed}=(63\pm 4)\%$ is
lower than the expected one, due to the experimental imperfections
and to the emission of higher number of photons by the first
crystal. In Fig.\ref{fig:injected_NOON} we report the oscillation
of the injected field with (Fig.\ref{fig:injected_NOON}-(b)) and without
(Fig.\ref{fig:injected_NOON}-(a)) the conditional generation of the two photon
NOON state by the first crystal. As shown, the two-fold
coincidences present a $\lambda$ period and the $\lambda/4$
feature is displayed only by the fourfold coincidences
Fig.\ref{fig:injected_NOON}-(b).

\begin{figure}[t]
\centering
\includegraphics[width=0.50 \textwidth]{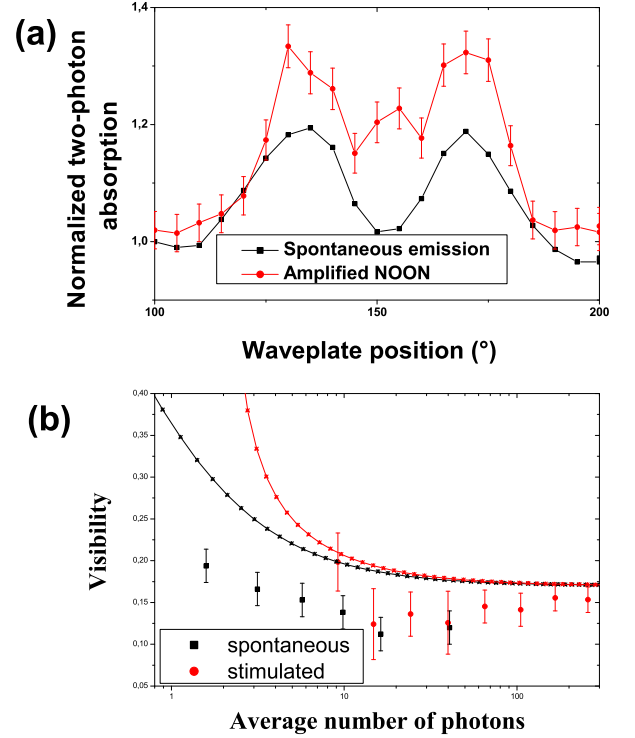}
\caption{(a) Oscillation fringe patterns in the stimulated and
spontaneous regimes. The unbalanced minima are due to a different
coupling of the $\vec{\pi}_{H}$ and $\vec{\pi}_{V}$ polarized
signals with the single mode fiber. (b) Visibility value
$\mathcal{V}^{(2)}_{N=2}$ as a function of NL gain g in the
spontaneous (triangular dots) and stimulated case (circular dots).
Experimental (points) and theoretical trend of visibility in the
stimulated regime (curve) are shown. The theoretical curve used are:
$V_{sp} = 0.85 \, \mathcal{V}^{(2)}_{N=0}$ and 
$V_{stim} = 0.85 \, \mathcal{V}^{(2)}_{N=2}$. The factor 0.85 has been inserted to 
consider experimental imperfections. The curves are parametric plotted as a
function of the respective number of generated photons, which are 
$\langle \hat{n} \rangle_{sp} = 2 \sinh^{2} g$ and 
$\langle \hat{n} \rangle_{stim} = 2 + 6 \sinh^{2} g$.
Data in the spontaneous regime refer to work \cite{DeMa08}.}
\label{fig:3}
\end{figure}

We characterized then the state generated by the second crystal by
evaluating the correlation function $\mathcal{G}^{(2)}$\ in the
spontaneous, by detecting coincidences between detectors
$\{D_{1}^{A},D_{1}^{B},D_{T}\}$, and stimulated regime, by
detecting coincidences between detectors
$\{D_{1}^{A},D_{1}^{B},D_{T},D_{T}^{*}\}$ ,\ for a value of the NL
gain $g=2$ \cite{Eise04,Cami06}.\ We observed that the $\lambda
/2$\ period has been preserved by the amplification process
Fig.\ref{fig:3}-(a).\ The  minima, at $105^{\circ}$ and
$150^{\circ}$, correspond to polarizations $\vec{\pi}_{H}$ and
$\vec{\pi}_{V}$. The unbalancing between the two values of the
absorption rate is due to a different coupling of the two
orthogonal linear polarizations with the single mode fiber. This
effect is related with the distinguishability, i.e spectral
difference, between the ordinary and extraordinary wave vectors
cones generated by the second crystal during the amplification
process. The visibility has been evaluated through the definition
$\mathcal{V}=\frac{C^{(4)}_{max}-C^{(4)}_{min}}{C^{(4)}_{max}+C^{(4)}_{min}}$,
where $C^{(4)}$ is the value of the fourfold coincidences. In
particular, only a portion of the global fringe pattern of
Fig.\ref{fig:3} has been used to calculate the visibility. Only
the maximum and the adiacent minimum which exhibit the higher
contrast were considered, as  a $\frac{\pi}{N}$ interval of
the fringe pattern, showing a $\frac{\lambda}{2N}$ resolution, is
necessary for quantum lithographic applications. By the same
measurement we observe the fringe patterns for different gain
values by increasing the UV pump beam. We report in Fig.
\ref{fig:3}-(b) the trend of visibility as a function of the
number of photons generated:\ the spontaneous visibilities have
been taken from \cite{DeMa08}. The experimental data are compared
with theoretical predictions in both regimes: spontaneous and
stimulated. The theoretical trends have been scaled by a factor
$0.85$, that was the asymptotical visibility obtained in the
spontaneous case in \cite{DeMa08}, due to experimental
imperfections. In the amplified case the experimental asymptotical
visibility is affected both by experimental imperfections and by
the emission of higher number of photons by the first crystal. We
observe that both the data points for increasing gain values move
away from the theoretical trends. This can be due to a partial
multimode operation of the parametric amplifier \cite{Thav04}.
We conclude that the value of visibility in the two regime is almost
the same, but an enhancement of the signal in the
stimulated case has been observed. Indeed the probability of
observing a sub-Rayleigh phenomenon is proportional to the second
order correlation function $\mathcal{G}^{(2)}$ in both regimes. By
the theory, the stimulated signal is seven time higher than  the
spontaneous one:
$\frac{\mathcal{G}_{N=2}^{(2)}}{\mathcal{G}_{N=0}^{(2)}}=7$.
Experimentally we can evaluate this ratio by the following method:
the probability of detecting coincidences in the spontaneous case
is $P_{sp}=\frac{C^{(2)}}{R}$, where $C^{(2)}$ are coincidences
between detectors $\{D_{1}^{A},D_{1}^{B}\}$ and $R$ is the
repetition rate. In the stimulated case it reads :
$P_{stim}=\frac{C^{(4)}}{\Xi}$, where $C^{(4)}$ are coincidences
between detectors $\{D_{T},D_{T}^{*},D_{1}^{A},D_{1}^{B}\}$, and
$\Xi$ are coincidences between detectors
$\{D_{T},D_{T}^{*}\}$ on trigger mode, that is the rate of injection of
the two photon NOON state in the QIOPA per second. Hence the ratio between the
two probabilities is : $\frac{P_{stim}}{P_{sp}}=(6.21\pm 0.8)$.

\subsection{Amplification of $N>2$ states}
As last step, we investigate the amplification of N-photon NOON states with N%
$>$2  with the same device. The injection of a 3-photon state $%
|\psi^{3} \rangle_{1} =2^{-1/2}\left( |3+\rangle -|3-\rangle
\right)_{1} $ leads to an amplified wave function of the form:
\begin{widetext}
\begin{equation}\label{eq:wf_coll_3}
\begin{aligned}
\ket{\Phi^{3}} &= \frac{1}{\sqrt{12} C^{4}} \sum_{i,j=0}^{\infty} \frac{\left( \frac{\Gamma}{2} \right)^{i} \left( - \frac{\Gamma}{2} \right)^{j}}{i! j!} \left\{ \sqrt{(2i+3)! 2j!} \ket{(2i+3)+,(2j)-} - \sqrt{2i! (2j+3)!} \ket{(2i)+,(2j+3)-} \right\} + \\
&- \frac{\Gamma \sqrt{3}}{2 C^{2}} \sum_{i,j=0}^{\infty} \frac{\left( \frac{\Gamma}{2} \right)^{i} \left( - \frac{\Gamma}{2} \right)^{j}}{i! j!} \left\{ \sqrt{(2i+1)! 2j!} \ket{(2i+1)+,(2j)-} + \sqrt{2i! (2j+1)!} \ket{(2i)+,(2j+1)-} \right\}
\end{aligned}
\end{equation}
\end{widetext}

Let us analyze the expression of the quantum state in equation
(\ref{eq:wf_coll_3}). We expect the third order correlation
function to have oscillation in all the three harmonics $\theta$,
$2 \theta$ and $3 \theta$. In fact, the first part of the wave
function contains the sum of quantum states of the form
$\ket{2i+3,2j} - \ket{2i, 2j+3}$. These are analogous to 3 photons
NOON states with a common background $2i,2j$ generated by the
crystal, thus leading to a $\frac{\lambda}{3}$ period. The same
argument holds for the second part of eq.(\ref{eq:wf_coll_3}), as
the unbalancement of only 1 photon determines a $\lambda$ period.
We finally expect the presence
of a $\frac{\lambda}{2}$ period due to the couple emission of photons by the crystal.\\
Explicit calculation of the third order correlation function gives
the result:

\begin{equation}
\mathcal{G}_{N=3}^{(3)} = a(\overline{n})+b(\overline{n})\cos(\theta)+c(\overline{n%
})\cos(2\theta)+d(\overline{n})\cos(3\theta)
\label{eq:third_order_abs_rate}
\end{equation}

\noindent where $a(\overline{n}) = 6 + 342\overline{n} + 1782 \overline{n}^{2} + 1824 \overline{n}^{3}$
 and $c(\overline{n}) = \frac{1}{2} \left[ 81 \overline{n} + 369 \overline{n}^{2} + 288 \overline{n}^{3} \right]$
 are third
degree polynomial in $\overline{n}$,\
while
$b(\overline{n}) = \frac{3}{2} \left[ 3 \overline{n}^{2} + 3 \overline{n} + 1 \right]$
and $d(\overline{n}) = - \frac{27}{2} \left[ \overline{n} + \overline{n}^{2} \right]$
are second degree polynomial in $\overline{n}$.
We find, as said, the presence of oscillating terms
at the three fundamental harmonics in $\theta $, $%
2\theta $ and $3\theta $.\ The term in $2\theta$ is dominant for
high gain values, and the intrinsic oscillation of the crystal
with period $\frac{\lambda }{2}$ suppresses the amplitude of the
oscillations with  $\frac{\lambda }{3}$ period. Hence this
apparatus based on the collinear QIOPA device cannot be used for
the amplification of a generic  state, as the interference pattern
of the seed is masked during the amplification process.\\
Hence in order to preserve the $\frac{\lambda}{2N}$ phase
oscillation after the amplification process, a different amplifier
device,  not containing an intrinsic phase oscillation,
has to be employed.

\section{Non collinear amplifier}
\label{sec:non_coll}
\begin{figure}

    \centering
        \includegraphics[width=0.45\textwidth]{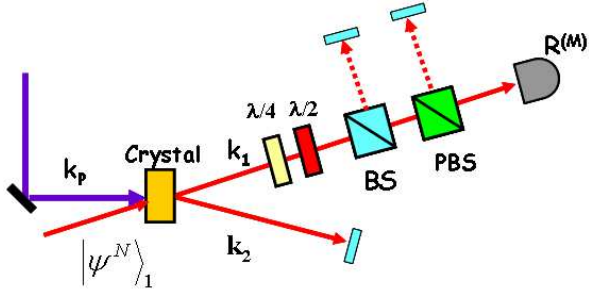}
    \caption{Experimental setup for the amplification of a NOON state
by a non-collinear amplifier, implemented by a type-II cut BBO
crystal in non collinear configuration. The state
$|\protect\psi^{N} \rangle_{1} $ is injected into the input mode
$\mathbf{k}_{1}$.
The BS is inserted in order to simulate losses.}
    \label{fig:schema_non_coll}
\end{figure}

In this section we study the amplification
of NOON states exploiting an Optical Parametric Amplifier working in a non-collinear configuration.
The interaction Hamiltonian of this device is \cite{DeMa05}:

\begin{equation}
\label{eq:int_Ham_non_coll}
\hat{\mathcal{H}}_{int} = \imath \hbar \chi \left( \hat{a}^{\dag}_{1\pi} \hat{a}^{\dag}_{2\pi_{\bot}} - \hat{a}^{\dag}_{1\pi_{\bot}} \hat{a}^{\dag}_{2\pi}\right) + \mathrm{h.c.}
\end{equation}
where $\pi, \pi_{\bot}$ stand for any two orthogonal polarizations,
as this configuration is invariant under SU(2) rotations.\\
The proposed scheme is shown in Fig.\ref{fig:schema_non_coll}.

After the preparation of the seed, the state is injected on mode
$\mathbf{k}_{1}$ in the amplifier together with the pump beam
$\mathbf{k}_{p}$ to obtain the amplification process. A phase
shift $\theta$ is then introduced between the two polarization
$\vec{\pi}_{+}, \vec{\pi}_{-}$ and the M-th order absorption
process is performed in $R^{(M)}$. An unbalanced BS with
transmittivity $\eta$ will be subsequently introduced in
Sec.\ref{sec:loss} to simulate losses and non unitary
efficiency of detection. This scheme corresponds to evaluating the
M-th order correlation defined by the operator $\hat{G}^{(M)} = [
\hat{c}_{1}^{\dag}(t) ]^{M} [ \hat{c}_{1}(t) ]^{M}$, where
$\hat{c}_{1}^{\dag}$ is the creation operator associated to the
revealed mode  correspondind to the Heisenberg evolution of the
field operator $\hat{a}_{1H}$:

\begin{equation}
\label{eq:reveal_mode_non_coll}
\hat{c}_{1}^{\dag}(t) = \frac{1}{\sqrt{2}} \left[ \hat{a}_{1+}^{\dag}(t) - e^{\imath \theta} \hat{a}_{1-}^{\dag}(t) \right]
\end{equation}

The time evolution of the field operators in the crystal is
derived from the interaction Hamiltonian of the non-collinear OPA
(\ref{eq:int_Ham_non_coll}). The Heisenberg equations gives:
\begin{eqnarray}
\hat{a}^{\dag}_{1+}(t) &=& \hat{a}^{\dag}_{1+} \cosh(g) + \hat{a}_{2-} \sinh(g)\\
\hat{a}^{\dag}_{1-}(t) &=& \hat{a}^{\dag}_{1-} \cosh(g) + \hat{a}_{2+} \sinh(g)
\end{eqnarray}
where $g = \chi t_{int}$ is the non-linear gain of the process.

\subsection{Spontaneous emission}\label{sec:spont_non_coll}
Let us study  the interferometrical feature of this device in
the spontaneous emission case. It will be shown that the spontaneous emitted field does not 
show any oscillation patterns for any orders of correlation.\\
The unitary time evolution operator in the interaction picture for the non-collinear OPA
can be written in the form \cite{DeMa05}:
\begin{equation}
\label{eq:U_non_coll}
\begin{split}
\hat{U} &= e^{\Gamma \left( \hat{a}^{\dag}_{1+} \hat{a}^{\dag}_{2-} - \hat{a}^{\dag}_{1-} \hat{a}^{\dag}_{2+} \right)} e^{-\ln C \left( 1 + \hat{n}_{1+} + \hat{n}_{1-} + \hat{n}_{2+} + \hat{n}_{2-} \right)} \\
& e^{\Gamma \left( \hat{a}_{1-} \hat{a}_{2+} - \hat{a}_{1+} \hat{a}_{2-} \right)}
\end{split}
\end{equation}
where $C = \cosh(g)$ and $\Gamma = \tanh(g)$. Applying this
operator to the input vacuum state we obtain:

\begin{equation}
\label{eq:phi_spont}
\ket{\Phi} = \frac{1}{C} \sum_{n=0}^{\infty} \Gamma^{n} \sum_{m=0}^{n} \ket{(n-m)+,m-}_{1} \ket{m+,(n-m)-}_{2}
\end{equation}

The M-th order correlation function, calculated in the Heisenberg
picture shows that there is no oscillation pattern in the
spontaneous radiation. Let us ignore for now the effects of
losses, and evaluate $\mathcal{G}^{(M)}_{0} = \bra{0}
\hat{G}^{(M)} \ket{0}$. The M-th order correlation operator reads:

\begin{equation}
\begin{split}
\hat{G}^{(M)} &= \frac{1}{2^{M}} \left[ \hat{a}^{\dag}_{1+} C + \hat{a}_{2-} S - e^{\imath \theta} \hat{a}_{1-}^{\dag} C - e^{\imath \theta} \hat{a}_{2+} S \right]^{M} \times \\
&\times \left[ \hat{a}_{1+} C + \hat{a}^{\dag}_{2-} S - e^{- \imath \theta} \hat{a}_{1-} C - e^{- \imath \theta} \hat{a}^{\dag}_{2+} S \right]^{M}
\end{split}
\end{equation}

where $S = \sinh(g)$. This operator can be written, using the
multinomial expansion, as:

\begin{equation}
\label{eq:G_N_oper_expan}
\begin{aligned}
\hat{G}^{(M)} &= \frac{1}{2^{M}} \left( \sum_{i,j,k} g_{ijk} ( \hat{a}_{1+}^{\dag} )^{M-i-j-k} ( \hat{a}_{2-} )^{i} ( \hat{a}_{1-}^{\dag} )^{j} ( \hat{a}_{2+} )^{k} \right) \\
&\times\left( \sum_{l,m,n} g^{\ast}_{lmn} (\hat{a}_{1+}^{\dag} )^{M-i-j-k} ( \hat{a}_{2-} )^{i} ( \hat{a}_{1-}^{\dag} )^{j} ( \hat{a}_{2+} )^{k} \right)
\end{aligned}
\end{equation}

where:

\begin{equation}
g_{ijk} = \left( \frac{1}{\sqrt{2}} \right)^{M} \begin{pmatrix} M \\ i,j,k \end{pmatrix} (-1)^{j+k} \left( e^{\imath \theta} \right)^{j+k} C^{M-i-k} S^{i+k}
\end{equation}

and the sums are extended as $\sum_{i=0}^{M} \sum_{j=0}^{M-i}
\sum_{k=0}^{M-i-j}$.\\
The average of $\widehat{G}^{(M)}$ on the vacuum input state
gives:
\begin{equation}
\mathcal{G}^{(M)}_{0} =  M!\, S^{2M}
\end{equation}
This expression is independent on the phase for any order of the
correlation. Thus, no intrinsic phase oscillation pattern is
present in the radiation emitted in the spontaneous regime by the
non collinear OPA, as expected from the form of the interaction
Hamiltonian of eq.(\ref{eq:int_Ham_non_coll}).

\subsection{Amplified NOON quantum state}
First we calculate the quantum state
in the interaction picture. The amplified field is obtained, with a procedure completely
analogous to the spontaneous emission case calculated in
Sec.\ref{sec:spont_non_coll}, by applying the operator
(\ref{eq:U_non_coll}) to the injected state: $\ket{\Phi^{N}} =
\hat{U} \ket{\psi^{N}}_{1}$. The output state reads:

\begin{widetext}
\begin{equation}\label{eq:wf_non_coll_N}
\begin{aligned}
\ket{\Phi^{N}} &= \frac{1}{\sqrt{2} \sqrt{N!} C^{N+1}} \sum_{n=0}^{\infty} \Gamma^{n} \sum_{m=0}^{n} (-1)^{m} \times \\
&\times \left[ \sqrt{\frac{(n-m+N)!}{(n-m)!}} \ket{(n-m+N)+,m-}_{1} - \sqrt{\frac{(m+N)!}{m!}} \ket{(n-m)+,(m+N)-}_{1} \right] \ket{m+,(n-m)-}_{2}
\end{aligned}
\end{equation}
\end{widetext}

Let us analyze the expression (\ref{eq:wf_non_coll_N}): the
N-photons in excess on the two polarization modes with respect to
the spontaneous emission case of eq.(\ref{eq:phi_spont}) are
responsible for the $\frac{\lambda}{2N}$ fringe pattern. Hence the
original N photons in the injected state are  added to a
background field emitted by the crystal.

\subsection{M-th order correlation function}\label{sec:amplified}
We now calculate the generic M-th order correlation function
with the injection of a NOON state defined by the average
$\mathcal{G}^{(M)}_{N} = _{1}\bra{\psi^{N}} \hat{G}^{(M)}
\ket{\psi^{N}}_{1}$. It will be shown that the original features
of the injected seed will be maintained. It will be explicitly
demonstrated that the correlation functions of order $M<N$ do not
have any oscillation patterns, while the ones with $M \geq N$
exhibit sub-Rayleigh
$\frac{\lambda}{2N}$ feature. \\
The value of the correlation functions can be calculated in the
Heisenberg picture analogously to the spontaneous case of
Sec.\ref{sec:spont_non_coll}. We obtain the following expression
for $\mathcal{G}^{(M)}_{N}$:

\begin{equation}
\label{eq:GMN_M>N}
\begin{aligned}
\mathcal{G}^{(M)}_{N} &= \frac{M!}{2^{M}} \left\{\sum_{i=0}^{M-N} \sum_{j=0}^{N} C^{2j} S^{2(M-j)} \begin{pmatrix} N \\ j \end{pmatrix} \begin{pmatrix} M \\ i,j \end{pmatrix} \right.+\\
&+ \sum_{i=M-N+1}^{M} \sum_{j=0}^{M-i} C^{2j} S^{2(M-j)} \begin{pmatrix} N \\ j \end{pmatrix} \begin{pmatrix} M \\ i,j \end{pmatrix} + \\
& \left. - \left(-1\right)^{M} C^{2N} S^{2(M-N)} \left[ \sum_{i=0}^{M-N} \begin{pmatrix} M \\ i,N \end{pmatrix}\right] \cos(N \theta) \right\}
\end{aligned}
\end{equation}

for $M \geq N$, while for $M<N$ we obtain:

\begin{equation}
\label{eq:GMN_M<N}
\mathcal{G}^{(M)}_{N} = \frac{M!}{2^{M}} \sum_{i=0}^{M} \sum_{j=0}^{M-i} C^{2j} S^{2(M-j)} \begin{pmatrix} M \\ i,j \end{pmatrix} \begin{pmatrix} N \\ j \end{pmatrix}
\end{equation}

The form of eq.(\ref{eq:GMN_M>N}) explicitly shows the
$\frac{\lambda}{N}$ period of the emitted radiation, as only
constant or oscillating terms in $N \theta$ are present. 

\subsection{Losses and decoherence effects}
\label{sec:loss}

We are now interested in studying the effects of losses and of non
unitary efficiency of detection on the amplified field. We
introduce an unbalanced BS of transmittivity $\eta$ in spatial
mode $\mathbf{k}_{1}$ (Fig. \ref{fig:schema_non_coll}). The two BS
input modes are labelled by  $\hat{b}_{1}^{\dag}(t)$ and
$\hat{c}_{1}^{\dag}(t)$, where the second one is the OPA output
mode and the first one is the vacuum input lossy channel. The
revealed output mode corresponds to the field operator:
\begin{equation}
\label{eq:dec_operator}
\hat{d}_{1}^{\dag}(t) = \sqrt{\eta} \, \hat{c}^{\dag}_{1}(t) + \imath \sqrt{1 - \eta} \, \hat{b}^{\dag}_{1}(t)
\end{equation}
where the BS I/O relations have been used. The M-th order
correlation function $\widetilde{\mathcal{G}}^{(M)}_{N} = \, _{b_{1}}\bra{0}
_{c_{1}}\bra{\psi^{N}} \left\{ [\hat{d}_{1}^{\dag}(t)]^{M} [\hat{d}_{1}(t)]^{M} \right\}
\ket{\psi^{N}}_{c_{1}} \ket{0}_{b_{1}}$ reads:

\begin{equation}
\begin{aligned}
\widetilde{\mathcal{G}}^{(M)}_{N} &= \sum_{i,j=0}^{M} \left( \sqrt{\eta} \right)^{i+j} \left(\imath \sqrt{1 - \eta} \right)^{2M-i-j} \left( -1 \right)^{M-j} \times \\
& _{b_{1}}\bra{0} _{c_{1}}\bra{\psi^{N}} \left\{ [\hat{c}_{1}^{\dag}(t)]^{i} [\hat{b}_{1}^{\dag}]^{M-i} [\hat{c}_{1}(t)]^{j} [\hat{b}_{1}]^{M-j} \right\} \ket{\psi^{N}}_{c_{1}} \ket{0}_{b_{1}}
\end{aligned}
\end{equation}

The input vacuum field on mode $\hat{b}_{1}$ imposes the constraints $i=M$ and $j=M$ when we evaluate the average.
The correlation function $\widetilde{\mathcal{G}}^{(M)}_{N}$ then reads:

\begin{equation}
\label{eq:GMN_dec}
\widetilde{\mathcal{G}}^{(M)}_{N} = \eta^{M} \,_{c_{1}}\bra{\psi^{N}} [\hat{c}_{1}^{\dag}(t)]^{M} [\hat{c}_{1}(t)]^{M} \ket{\psi^{N}}_{c_{1}} = \eta^{M} \mathcal{G}^{(M)}_{N}
\end{equation}

Hence, the presence of losses and of the non unitary efficiency of
detection do not change the oscillation pattern of the amplified
field, but only reduce the efficiency of the process by a factor
$\eta^{N}$. The main difference between the pure NOON state and
the amplified field is in their resilience to losses. For the
injected state, as explained in Sec.\ref{sec:seed}, the loss of
just a single photon cancels the $\frac{\lambda}{N}$ behaviour of
the field and only a fraction $\eta^{N}$ contributes to the
successful events rate. On the contrary, for the amplified field
the non-linear gain of the process can be chosen so that $\eta^{M}
\overline{n} \gg 1$. In this condition, the large majority of the
pulses give contribution to the M-th order correlation and the
successful events rate is substantially not reduced.

\subsection{Asymptotical visibilities}
Knowing the correlation function, we can calculate the
visibilities associated to this M-photon absorption processes. We
can see by the form of eq.(\ref{eq:GMN_dec}) that the visibility
associated to the M-th order correlation function is not affected
by losses and by non unitary detection efficiency. Restricting our
attentions to the asymptotical visibilities, corresponding to $g
\rightarrow \infty$ and hence to an ideal infinite number of
photons in the emitted field, we obtain for a NOON state with
N=2,3,4 the following expressions:

\begin{eqnarray}
\widetilde{\mathcal{V}}_{N=2}^{(M)}(\overline{n} &\rightarrow &\infty )=\frac{M^{2}-M}{%
M^{2}+7M+8} \\
\widetilde{\mathcal{V}}_{N=3}^{(M)}(\overline{n} &\rightarrow &\infty )=\frac{%
M^{3}-3M^{2}+2M}{M^{3}+15M^{2}+56M+48} \\
\widetilde{\mathcal{V}}_{N=4}^{(M)}(\overline{n} &\rightarrow &\infty )=\frac{%
M^{4}-6M^{3}+11M^{2}-6M}{M^{4}+26M^{3}+203M^{2}+538M+384}
\end{eqnarray}

\begin{figure}[ht!]
    \centering
        \includegraphics[width=0.35\textwidth]{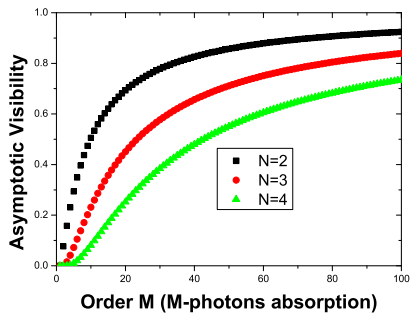}
    \caption{Plot of the  asymptotic ($g \rightarrow \infty $ and $%
\overline{n} \rightarrow \infty$) M-th order correlation function
as a function of the order M in three cases. The square data
corresponds to the injection of a 2-photons  state, the circular
data to a 3-photons state and the triangular data to a 4-photons
 state.}
    \label{fig:Mth_order}
\end{figure}

The plot of these three functions are reported in Fig.\ref{fig:Mth_order}.
We observe that the values of the visibilities grow with the order
of correlation M and decrease as the number of photons of the
injected  states increases. This is due to the characteristic of
the N-photon NOON seed, which implies an increase of both the
minimum and the maximum of the fringe pattern proportional to the
number of photons N.

\section{conclusion}
In this paper we investigated the amplification of a two photons
NOON state using two different scheme both based on the process of
optical parametric amplification. In Sec.\ref{sec:seed} we
reviewed how this kind of quantum states are extremely sensitive
to losses. In Sec.\ref{sec:exp} we propose to use a collinear
optical parametric parametric amplifier to amplify a 2-photon
entangled state, maintaining in the output field the
$\frac{\lambda}{4}$ pattern of the injected seed. We analyzed the
problem theoretically and experimentally , comparing the amplified
with the spontaneous emission regime analyzed in \cite{DeMa08}. We
found experimentally that the two regimes have comparable
visibilities, while the advantage of the stimulated case is a
significant increase of the number of photons in the emitted
radiation. We then showed that this device, due to the intrinsic
$\frac{\lambda}{4}$ oscillation of the radiation emitted by the
crystal, cannot be used to amplify a generic N photon states. We
then propose in Sec.\ref{sec:non_coll} to use a non-collinear
optical parametric amplifier to amplify a generic NOON state. We
showed that the oscillation period of the seed is maintained
during the amplification process and the visibility reaches an
asymptotical unitary value when the M-th order correlation
function with sufficiently high value of M is analyzed.
Furthermore, we showed that the amplified field exhibits a high
resilience to losses with respect to the extreme sensitivity of
the NOON states.

We acknowledge useful and stimulating discussions with Jonathan P. Dowling. This work was supported by the PRIN 2005 of MIUR and Progetto
Innesco 2006 (CNISM).

\end{document}